\documentclass[onecolumn,showpacs,preprintnumbers,amsmath,amssymb]{revtex4}
\preprint{}
\usepackage{graphicx}% Include figure files
\usepackage{dcolumn}% Align table columns on decimal point
\usepackage{bm}% bold math
\usepackage{mathrsfs}
\begin{document}
\title{Secure communication with choice of measurement}
\author{Dong  Xie}
\email{xiedong@mail.ustc.edu.cn}
\author{An Min Wang}
 \email{anmwang@ustc.edu.cn}
  \affiliation{Department of Modern Physics , University of Science and Technology of China, Hefei, Anhui, China.}
\begin{abstract}
It has been found that the signal can be encoded in the choice of the
measurement basis of one of the communicating parties, while the outcomes of the measurement are
irrelevant for the communication and therefore may be discarded. The proposed protocol was novel and interesting, but it wasn't secure for communication. The eavesdropper can obtain the information without being detected. We utilize the outcomes of the measurement and separate the Hilbert space to  propose a secure communication protocol. And the error correction code is used to increase the fault tolerance in the noise.

\end{abstract}

\maketitle

\section{Introduction}
Quantum measurement must have an unavoidable disturbance to the system. Recently, Amir Kalev at al. \cite{lab1} showed that this disturbance can be used for communication.
Nonselective measurements on one part of a quantum system in a maximally entangled state are used to encode and communicate information, where the outcomes are not recorded. Obviously, in the classical theory, the nonselective measurements cannot carry information \cite{lab2,lab3}. Hence, the process of quantum measurement can decompose into two stages: the first stage is the nonselective measurement, which elevates a particular set of dynamical
variables (those labeling the basis in our case) to reality; the second stage involves the determination of the value of the dynamical variable (is same with the classical measurement).

Compared with classical communication, the biggest advantage of quantum communication is the unconditional security. However the proposed protocol in the ref. \cite{lab1} is easily eavesdropped. So it is necessary and interesting to find a secure protocol for communication.
We find that although the outcomes of measurements on one part  are
irrelevant for the communication, they can be used to guarantee the security.

The noise from environment inevitably disturbs the quantum system. The general damage on the system comes from the decoherence induced by the noise. Our protocol can filter the incorrect information which are created by the decoherence noise. For those noise which can change the phase and the message, repetition code is used to correct the error.

The rest of article is arranged as follows. In the section II, we make a simple review and correction about the nonselective measurement protocol. A secure communication protocol is proposed in the section III. In the section IV, We advise the communication protocol in consider of the noise. Finally, we make a conclusion and outlook in the section V.

\section{A Simple Review of nonselective  measurement}
In order to introduce a secure communication protocol, we firstly review the nonselective  measurement protocol in the ref. \cite{lab1}.

There are $d+1$ mutual unbiased bases, which are given in terms of the computational basis by
\begin{equation}
|m;b\rangle=\frac{1}{\sqrt{d}}\sum_{n=0}^{d-1}|n\rangle w^{bn^2-2nm};b,m=\ddot{0},0,1,...,d-1,
\end{equation}
where $|m;\ddot{0}\rangle=|m\rangle$, $w=e^{i2\pi/d}$, and $d$ is an odd prime number.
Let Alice prepare one of the following $d^3$  two-qudit maximally entangled states \cite{lab4,lab5}:
\begin{equation}
|c,r;s\rangle_{1,2}=\frac{1}{\sqrt{d}}\sum_{n=0}^{d-1}|n\rangle_1|c-n\rangle_2w^{sn^2-2rn},
\end{equation}
in which, $c, r, s=0,1,...,d-1$. Then Alice sends one qudit labeled by 1 to Bob.
To communicate a message to Alice, Bob measures his qudit in the basis $|m; b\rangle$. The different $b$ is equivalent to different signal. After the measurement, Bob sends the qudit back to Alice. In order to get the message, Alice measures the two qudits in the basis of preparation, $\{{|c',r';s\rangle_{1,2}}\}_{c', r'=0}^{d-1} $ of Eq.(2). Finally, the decoding table is given by
\begin{equation}
\begin{split}
c\neq c'&\rightarrow b=s+\frac{r-r'}{c'-c},\\
r\neq r', c=c'&\rightarrow b= \ddot{0},\\
r=r', c=c'&\rightarrow inconclusive.
\end{split}
\end{equation}
The probability of the inconclusive outcome is $1/d$. So for large dimension $d$, Bob can communicate message to Alice very well. From the Eq.(3), the value of $b$ can be outside the range of $0,1,2...,d-1$. We make a correction as follows,
\begin{equation}
c\neq c'\rightarrow b=s+\frac{r-r'+Nd}{c'-c},
\end{equation}
where $N=0,\pm1,\pm2,...,\pm \infty$. The value $N$ is chosen to guarantee the value of $b$ in the range of $1,2...,d-1$.

\section{Secure communication protocol}
The above protocol isn't safe. The Eve can steal the information without being detected. The security is most important in the communication \cite{lab6}.

Eve can eavesdrop by the following way, as shown in Fig.1. Firstly, Eve intercepts the qudit $1$ which is sent from Alice to Bob. And Eve sends one of the $d^3$ two-qudit maximally entangled state, the one labeled by 3 to Bob. Next, Eve intercepts the qudit, which is sent from Bob. After Eve measures the two-qudit in the basis $\{{|c',r';s\rangle_{3,4}}\}_{c', r'=0}^{d-1} $, she can obtain the message, the value of $b$. Then she measures the the qudit $1$ in the basis $|m,b\rangle_{m=0}^{d-1}$. So Alice and Bob cannot know that the communication has been eavesdropped, because Alice also obtains the corrected message from Bob.
\begin{figure}[h]
\includegraphics[scale=0.20]{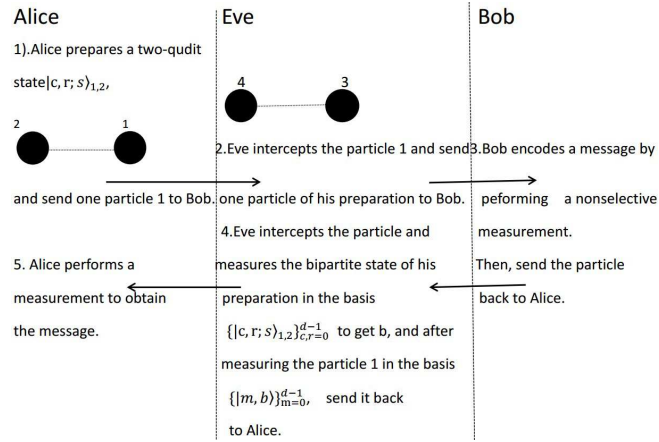}
\caption{\label{fig.1}The eavesdropping scheme of Eve in the communication. }
 \end{figure}

 In order to guarantee the safety, it is necessary to detect Eve. We consider the dimension of the Hilbert space to be $\frac{3d-1}{2}$. So the computational basis is $\{|n\rangle\}_{n=0}^{\frac{3d
 -3}{2}}$. The kets that compose other bases (which we care about) are given by
 \begin{equation}
 \begin{split}
 &|m,x;b\rangle=\frac{1}{\sqrt{d}}\sum_{n=\frac{x-1}{2}(d-1)}^{\frac{x+1}{2}(d-1)}|n\rangle w^{bn^2-2nm};\\
 &b,m=0,1 ..., d-1; x=1,2;\\
 &|m,x;b=\ddot{0}\rangle=|m,x\rangle.
 \end{split}
 \end{equation}
 Where $x$ represents the different Hilbert space and $w=e^{i2\pi/d}$.

 Alice prepares one of the following two-qubit maximally entangled states in the space $x$:
 \begin{equation}
 \begin{split}
 &|c,r,x;s\rangle_{1,2}=\frac{1}{\sqrt{d}}\sum_{n=\frac{x-1}{2}(d-1)}^{\frac{x+1}{2}(d-1)}|n\rangle_1|f(c-n)\rangle_2w^{sn^2-2rn},\\
 &c,r,s=0,1,...,d-1,
\end{split}
 \end{equation}
 in which,
  \begin{equation}f(c-n)=
 \begin{cases} c-n & c\geq n,\\c-n+\frac{x+1}{2}(d-1)+1 & c<n.
 \end{cases}
 \end{equation}
And sends the particle 1 to  Bob. Bob accepts and performs a measurement on the qudit in the basis
$\{|m,x,b\rangle_1\}_{m=0,x=1}^{d-1, 2}$. The different $b$ represents different signal. The two-qudits state is described now by
\begin{equation}
\begin{split}
\rho_{1,2}=&\sum_{m=0,x'=1}^{d-1,2}|m,x',b\rangle_1\langle m,x',b|\\
&|c,r,x;s\rangle_{1,2}\langle c,r,x;s||m,x',b\rangle_1\langle m,x',b|
\end{split}
\end{equation}
The proposed protocol in the ref. \cite{lab1}, Bob need not record the measurement outcome. For the security, the measurement outcome of $x$ is useful. After the measurement, Alice and Bob check the value of $x$. If the value of $x$ recorded by Bob is different with that from Alice, they abort the protocol; if the value of $x$ are same, Bob sends the particle 1 back to Alice. Finally Alice performs a measurement on the two-qudit in the basis, $\{|c',r',x;s\rangle_{1,2}\}_{c',r'=0}^{d-1}$, to retrieve the message from Bob. The decoding table is same as the Eq.(3).

In the process of communication, there are only two ways by which Eve can eavesdrop on the communication: measure or not measure the particle 1 before Bob to get the value of $x$.

First way, Eve steals message as the way in Fig.1. However, Eve doesn't know the right value of $x$ in the state sent from Alice. If she only chooses a value of $x$ randomly from $\{1, 2\}$, not to measure the particle 1 before Bob, she would be detected with the probability $3/4$ for one message by checking the value of $x$ between Alice and Bob. The probability of successful eavesdropping will decrease exponentially with the number of messages $N$: $(3/4)^N$. And, if one chooses a greater Hilbert space (the value of $x$ has more choices), the probability is very small for every message.

Second way, in order to know the value of $x$, Eve needs to perform a measurement on the particle 1. Due to the two subspaces separated by the variable $x$ are nonorthogonal, Eve could obtain the right value $x$ with the probability $1/2$ by the measurement without disturbing the state of particle 1. For $N$ messages, the probability is given by $(\frac{1}{2})^N$. Perhaps Eve can get the right value $x$ by making the same measurement on particle 1 as Bob. Then, she sends the particle 1 to Bob. However, Eve doesn't know the measurement of Bob beforehand, so the successful probability is $\frac{1}{d+1}$ for one message.

No matter which way Eve would use, the probability of successful eavesdropping tends to zero for many messages. Namely, the proposed protocol is secure against the eavesdropping.

The whole secure communication protocol can be simply summarized as follows.\\
1. Alice randomly chooses one two-qudit state from Eq.(6) and send particle 1 to Bob.\\
2. Bob accepts the particle 1, measures it in the basis $\{|m,x,b\rangle_1\}_{m=0,x=1}^{d-1, 2}$, and records the value of $x$.\\
3. After Bob obtains the value of $x$, Alice tells Bob the initial value of $x$ by classical communication. If the value of $x$ from Alice and Bob are different, they abort the communication.\\
4. Bob sends the particle 1 back to Alice. Then Alice performs a measurement on the two-qubit and retrieves the message $b$.\\
5. Bob sends $2n$ messages to Alice by the above procedure.\\
6. Alice randomly selects $n$ messages to check with Bob. They keep the remaining $n$ messages. If more than an acceptable number disagree, they abort the protocol.\\
7. Alice and Bob perform information reconciliation and privacy amplification on the remaining $n$ messages to obtain $l$ shared message \cite{lab7,lab8}.

\section{Secure protocol in the noise}
In the practical communication, the noise from the environment is inevitable. For secure and successful communication, we have to consider the influence of noise.  The noise destroys the quantum state and reduces the fidelity. In order to increase the fault tolerance, the error correction code is necessary. At the same times, we have to avoid treating the Eve as the noise, which is untractable.

Firstly, we consider that the noise only induces decoherence of quantum state. Decoherence is the most obstacle in quantum communication and computation \cite{lab7,lab9,lab10}.

We find that the function of decoherence on the maximally entangled state(Eq.(6)) is same like the measurement of Bob on the particle 1 in the basis, $\{|m,x;b=\ddot{0}\rangle_1\}_{m=0,x=1}^{d-1,2}$. So Alice will get the message $b=\ddot{0}$, no matter what Bob has sent. In order to find the error created by the decoherence noise, Bob never sends the message $b=\ddot{0}$. Once Alice obtains the message $b=\ddot{0}$, he immediately knows there is an error from the decoherence. Here, simple repetition code for correcting the error is easily utilized by Eve. Because Eve can also perform a measurement on the particle 1 in the basis, $\{|m,x;b=\ddot{0}\rangle_1\}_{m=0,x=1}^{d-1,2}$. We cannot distinguish her from noise. So, Eve just measure a few of repeated messages to get the value of $x$, and sends one particle of his preparation to Bob. Like the scheme in Fig.1, Eve can steal the message successfully.

There is one way which can simply solve the question in the decoherence noise. It isn't necessary to use the repetition code to correct the error from the decoerence noise. After the quantum communication, Alice tells Bob which messages are wrong, and discards them. The remaining messages are treated as secret keys.

Then, for the robust against the decoherence noise, the above way  performs very well. However, there are some noise, which can change the phase of quantum state to create the wrong message $b$. The above way isn't effective, because Alice don't know which messages are incorrect until Bob tells him. This makes the question become very hard. We proposed a scheme which can correct the error as follows.

Now repetition code is necessary to be used to find the error created by the noise which can change the phase and the message. A secure protocol is proposed as follows.\\
 The first step: Bob sends $2n*m$ messages to Alice, where we use repeated m messages to denote one message for correcting the error. The same $m$ messages needn't to be together, and are randomly arranged in the sequence of transmission as shown in Fig.2. \\
 The second step: After Alice obtains $2n*m$ messages, they compare $n*m$ messages which are selected randomly. If more than t of these disagree, they abort the protocol. Here, the value of $t$ is required to be chosen suitably. If $t$ is too small, the Eve can steal too much information; if $t$ is too big, the success probability of protocol is too low (we leave an open question how to choose the optimal value of $t$ ). \\
 The third step: Bob tells Alice the position of repeated message in the sequence.  Then Alice corrects the error according to the principle that the minority is subordinate to the majority.
 \begin{figure}[h]
\includegraphics[scale=0.20]{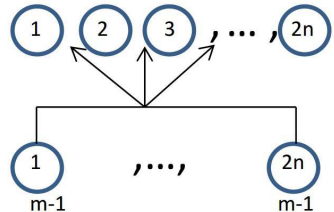}
\caption{\label{fig.2}Bob sends $2n*m$ messages by this diagram. Arrange 2n*(m-1)messages into the sequence of 2n messages randomly. }
 \end{figure}

\section{Conclusion and outlook}
We have proposed a secure communication protocol based on treating the choice of measurement as signal. It is suitable for some interesting cryptography tasks, such as the superdense coding
\cite{lab11} which achieves $2\log2d$ bits per qudit sent from Bob to Alice. In this article, the first stage of quantum measurement is used for communication and the second stage are utilized to guaranteed the security. We guess that a secure communication protocol must involves the second stages because the classical information is required to detect the eavesdropper Eve (a formal proof is left as an open question). And there are other secure communication protocols, which deserve the further study.

The noise also is considered. The error created by the decoherence noise is easily found, so it just need to discard the wrong messages. In general noise, a careful check is required. And we advise the repetition code to correct the error, where the sequence is randomly arranged against the eavesdropping.

\section{Acknowledgement}
This work was supported by the National Natural Science Foundation of China under Grant No. 10975125 and No. 11375168.

 \end{document}